\documentclass[epj]{svjour}
\RequirePackage{amsmath}
\RequirePackage[english]{babel}
\RequirePackage{braket} 
\RequirePackage[usenames, dvipsnames]{color}
\RequirePackage{amsfonts}
\RequirePackage{amssymb}
\RequirePackage{mathtools}
\RequirePackage{graphicx}

\journalname{Eur. Phys. J. C}

\begin{document}

\title{Diffusion coefficients and constraints on hadronic inhomogeneities in the early universe.}


\author{Sovan Sau 
       \and
       Sayantan Bhattacharya\thanks{\emph{Present Address:} University of Massachusetts Lowell, MA} 
\and Soma Sanyal}




\institute{University of Hyderabad
 Prof.C.R.Rao Road, Hyderabad, India 500046 \\
\email{sovan.sau@gmail.com,  sayantan34@gmail.com, sossp.uoh@nic.in}}


\date{Received: date / Accepted: date}

\abstract{Hadronic inhomogeneities are formed after the quark hadron phase transition. The nature of the phase transition 
dictates the nature of the inhomogeneities formed. Recently some scenarios of inhomogeneities have been discussed where the strange quarks 
are in excess over the up and down quarks.  The hadronization of these quarks will give rise to a large density of hyperons and kaons
in addition to the protons and neutrons which are formed after the phase transition. 
These unstable hyperons decay into pions, muons and their respective neutrinos. 
Hence the plasma during this period consists of neutrons, protons, electrons, muons and neutrinos. Due to the decay of the 
hyperons, the muon component of the inhomogeneities will be very high. We study the diffusion of neutrons and protons in the presence of a large number of muons 
immediately after the quark hadron phase transition. We find that the presence of the muons enhances the diffusion 
coefficient of the neutrons/protons. As the diffusion coefficient is enhanced, the inhomogeneities 
will decay faster in the regions where the muon density is higher. Hence smaller  muon rich inhomogeneities will be completely wiped out. 
The decay of the hyperons will also generate muon neutrinos. Since the big bang nucleosynthesis provides  constraints on the neutrino degeneracies, we revisit the effect of non zero degeneracies on the primordial elements.}


 
\maketitle

\section{Introduction}

The quark hadron phase transition in the early universe resulted in the formation of hadrons at around $200$ MeV. Before the
phase transition, the baryon number was carried by the nearly massless quarks, while after the phase transition the baryon
number is carried by the heavier hadrons. It has been proposed that if the phase transition is a first order phase transition, 
baryon number gets concentrated in between the bubble walls and baryonic inhomogeneities are formed at the end of the phase transition 
\cite{inhomo}. These baryon inhomogeneities affect the standard nucleosynthesis calculations. Later, lattice studies 
seemed to indicate that the phase transition is either a second order or a cross-over. However, even if the phase transition is not 
a first order transition, there is still the possibility of trapping a higher density of quarks in different 
regions and generating baryon inhomogeneities. There are scenarios where collapsing $Z(3)$ domain walls generate
inhomogeneities\cite{layek}. Dense inhomogeneities result in metastable quark nuggets \cite{quarknuggets}.
Metastable H dibaryons have also been predicted due to the presence of s-quarks at the time of the 
quark hadron transition \cite{dibaryons}. Recently some novel scenarios have been discussed which have reopened the case for the first order phase transition in the early universe \cite{boeckel,wygas}. It is well established that any first order QCD phase transition will result in baryon inhomogeneities in some form. So it seems that there is a strong possibility of an inhomogeneous distribution of baryon number across the 
universe after the quark hadron phase transition.

As the temperature cools down, neutrons and protons from these overdense region gradually diffuse to the underdense regions. 
The diffusion of neutrons and protons through the overdensities have been discussed previously in ref.\cite{ahs,bchitre,suh}.
When the diffusion of the neutrons and protons are 
studied, the standard interactions considered are the interactions between the neutrons, protons and electrons. The reason 
being that the other hadrons formed would decay in a short span of time. Even though the muon is an important part of the plasma 
at that temperature,  none of these calculations considered the collision of the neutrons (or protons) with the muons.
In this article, we argue that there are certain scenarios, where the muon density in the plasma cannot be neglected. 
If we look at the overdensities formed by the 
collapsing $Z(3)$ domains, we notice that a larger number of strange quarks are trapped in the overdense region as compared to the up and down 
quarks \cite{atreya}. This means there would be a large production of hyperons and kaons immediately after the phase transition. The hyperons 
would decay almost immediately but would result in the formation of a large number of pions and muons. Pions would also subsequently decay into muons.  Detailed discussion on the evaporation of quark nuggets formed in various cosmological scenarios also indicate that they mostly decay by the emission of kaons \cite{kaons}. 
Kaons themselves have a short lifetime and would subsequently decay into muons. There have also been discussions of antimatter domains 
formed after the quark hadron transition which produce a large number of pions \cite{antimatter}. These pions will also decay 
into muons. We therefore feel that there is every possibility that the baryon overdensities formed immediately after the quark hadron transition 
would have a significant number of muons too. Thus the diffusion coefficient of the protons and neutrons should also include their 
interaction with the muons. We would like to emphasize that this will be applicable mostly to 
the diffusion coefficients immediately after the quark hadron phase transition.  A recent study \cite{husdal} has shown that during this time, the inclusion of muons will increase the bulk viscosity roughly by a $100$ million times. Similarly, there have been studies of pions generated immediately after the quark hadron transition \cite{pionic} and their effects on the total entropy of the universe. 
Therefore  it is important to study the decay of hadronic inhomogeneities in the presence of a large number of muons. 

In this work we find the diffusion coefficient of the nucleons in the presence of muons. Generally at temperatures above $1$ MeV, neutrons and protons 
are in equilibrium with respect to weak interactions. Most of the studies of 
diffusive segregation of neutrons and protons are at temperatures below $1$ MeV. At this temperature, the weak interactions fall out of equilibrium and 
the neutron being neutral  moves faster than the proton. We are however interested in temperatures higher than $100$ MeV. Here the neutrons and protons change continuously into one another through weak interactions so the particles are treated as indistinguishable. 
 Our primary premise is that the number density of muons in certain regions with baryon over densities will be higher immediately after the phase transition.
We find the nucleon - muon scattering cross section in the temperature range of 100 MeV. This gives us the diffusion coefficient of the nucleons in these temperature ranges. 
We then use the diffusion coefficient to study the decay of inhomogeneities in the era after the quark hadron phase transition.

The other fall out of the decay of the unstable particles is the production of a large amount of muon neutrinos. This changes the muon neutrino chemical potential. Neutrino degeneracy and its effect on nucleosynthesis has been studied before \cite{olive}. Constraints on antimatter domains and other baryon inhomogeneities have also been obtained from 
nucleosynthesis calculations \cite{antimatter}. We revisit some of these calculations for the case of inhomogeneities which have a pre-dominance of strange quarks. As mentioned before, such inhomogeneities would form from the collapse of Z(3) domains around the time of the quark hadron transition.
Though, it is difficult to draw stringent constraints from the nucleosynthesis results due to the fact that it is a combination of all three neutrino 
degeneracies, one can still put some bounds on the muon neutrino degeneracy. We use one of the available nucleosynthesis codes based on the Wagoner - Kawano code \cite{wagoner} and modified 
by S. Dodelson \cite{dodelson} to look for constraints coming from nucleosynthesis. The nucleosynthesis code allows us to change the neutrino degeneracy parameters and obtain the abundances of the primordial elements. There have been previous studies of the effect of neutrino degeneracies on nucleosynthesis \cite{olive}. These are over a very wide range of baryon to photon ratios. We confine ourselves to the current value of the baryon to photon ratio and obtain the primordial abundances for different values of the chemical potentials for muon neutrino ($ \xi_\mu$) and the electron neutrinos ($\xi_e$).   

Since our basic starting point is the baryon overdense regions,
in the next section, we briefly discuss the formation of baryon overdensities and derive the diffusion coefficients of the nucleons through a plasma consisting of neutrons, protons, electrons and muons. In section 3, we follow it up by calculating the diffusion coefficients numerically between the temperature $200$ MeV - $100$ MeV. In section 4, we discuss the decay of the baryon inhomogeneities in a muon rich plasma  and compare it to the decay of the baryon inhomogeneity in a plasma which have equal number densities of electrons and muons. In section 5, we discuss the effect of muon neutrinos and obtain constraints on the inhomogeneities based on the  muon neutrino degeneracy parameters. Finally in section 6, we present our conclusions and some brief discussions.

\section{Baryon overdensities and diffusion coefficients}

\subsection{Generation of Baryon overdensities}

Baryon overdensities may be formed in the early universe during the quark
hadron phase transition. Initially they were formed during a first order 
phase transition.  A first order phase transition takes place with the nucleation of bubbles of the hadronic phase. As these bubbles move towards one another and coalesce, the baryon number gets concentrated in the small regions between the bubble walls. This happens as the baryon number prefers to be in the quark phase rather than the hadronic phase. Details of such baryon inhomogeneity formation can be obtained from ref.\cite{inhomo} and references therein.  Consequently lattice results indicated that the quark 
hadron phase transition may not be a first order phase transition. 
However, there are several situations which can arise in the early universe under which the QCD phase transition is still a first order phase transition \cite{wygas}. In such cases baryon inhomogeneities will form.  Baryon overdensities can also be formed during the electroweak phase transition \cite{megevand}. There are other ways in which such overdensities can be generated prior to the quark hadron transition. We are interested to look in detail at one such model where the baryon overdensity is generated by 
collapsing $Z(3)$ domain walls \cite{layek}. 

Layek et. al have discussed the generation of baryon overdensities by moving Z(3) domain walls. The profile of the overdensity is measured by $n(R)$ which is the baryon density left behind at a distance $R$ from the centre of the collapsing $Z(3)$ domain wall. This $n(R)$ can be about $1000$ times the background density for an area of radius $10$m. They had explicitly calculated the transmission coefficients of the $u,d,s$ quarks through the domain wall. They found that the number density of strange quarks is larger by an order of magnitude for the same size of the overdensity. So for  $R<1$, $n(R)$ for $u,d$ quarks is about $20,000$ while for $s$  quarks it is $6 \times 10^5$. Even if the parameters of the model are varied to generate lower overdensities, it has been found for the same radius, if $n(R)$ for $u,d$ quarks is about $400$, it is $5000$ for $s$  quarks.  
Hence these overdensities generated by the $Z(3)$ domain walls are dominated by strange quarks. Some of these overdensities may satisfy the conditions to form stable quark nuggets. However, since there are stringent conditions which need to be satisfied, most of these overdensities  will subsequently hadronize when the phase transition temperature is reached.

There are thermodynamical models which model the hadronization of 
quarks into hadrons \cite{thermo}. A detailed over-view of this hadronization process is given in ref. \cite{rafelski}. We mention only a few points about these models which are important for the calculation of number densities for the diffusion coefficients.

There are two phases here, one phase has the  $u,d,s$ quarks while the other phase consists of the hadrons, which are composites of the three quarks.  Since the universe is in thermal equilibrium, we are able to specify the grand canonical  partition function of the composite particles. The composite particles are the hadrons. The grand canonical  partition function depends on the temperature and the chemical potential of the particles. As the phase transition from the QGP phase to the HG phase takes about $10 \mu s$, chemical equilibrium is firmly established at the end of the phase transition. This enables us to associate the number density of the formed particles from the initial number density of the quarks. This has been discussed in detail in ref. \cite{rafelski}. They have shown the evolution of the chemical potential of the quarks in the absence of  inhomogeneities as well as the hadrons and mesons after hadronization. 
According to them the number density of the protons, neutrons, kaons and lambdas are of the order of $10^{35}$ particles/$cm^3$ around $100$ MeV. The kaons and hyperons formed will however be unstable and decay to pions and muons. It was shown in ref \cite{rafelski} that their number densities decrease to about $10^{20}$ particles/$cm^3$ at around $10$ MeV. We now see how these numbers will get affected due to the presence of the baryon inhomogeneities generated by the collapsing domain walls. 

The collapsing domain walls generate inhomogeneities where the number density of strange quarks is $10$ times greater than the magnitude of the 
up and down quarks. This means that $n_s \approx 10 n_d$. Now as is mentioned in ref.\cite{rafelski}, in chemical equilibrium, the chemical potential of hadrons is equal to the sum of the chemical potentials of their constituent quarks.  Since the Universe will still have to maintain the various constraints such as charge neutrality, constant entropy to baryon ratio etc, one will see a change in the number densities of the hadrons formed.
The number density of the particles is given by,
\begin{multline}
n_i = \frac{g_i}{2 \pi^2} \int_{m_i}^\infty dE E \sqrt{E^2 - m_i^2} \\ \times  \left( \frac{1}{e^{(E_i - \mu_i)/T} \pm 1}   - \frac{1}{e^{(E_i + \mu_i)/T} \pm 1} \right)
\label{rho}
\end{multline}
However, at temperatures close to the QCD phase transition, it has been argued  \cite{stuke} that the net number densities can be approximated to 
\begin{equation}
n_i = \frac{1}{6} g_i T^2 \mu_i + O(\mu_i^3)
\label{numberdensity} 
\end{equation}
for fermions. Since we are doing order of magnitude estimates here, we use this approximation to obtain an approximate value for the chemical potential $\mu_i$. The chemical potential is required as an input to the nucleosynthesis code. We calculate it from the number density using eq.\ref{numberdensity} later on in the fifth section. 

 The hadrons formed are the protons. neutrons, mesons and the hyperons. Since the number density of strange quarks are higher than the other quarks, a larger number of hyperons and kaons will be formed in the region of the inhomogeneities. 

Some of the particles like the sigma particle which have a very short lifetime ($(7.4 \pm 0.7) \times 10^{-20} $ secs) will instantaneously decay when formed. Since the hadronization occurs around $200 MeV$, the typical timescales are much longer. The Hubble time at the QCD phase transition is of the order of $10^{-5}$ secs \cite{schwarz}.  One thing that has been pointed out before, is that due to the background gas of photons and leptons, the timescale of the hadronic decay processes may not be the same as in vacuum \cite{rafelski}. The lambda particle with its mass closest to the protons and neutrons has a longer lifetime ($2.60 \times 10^{-10} $ s). This will decay into neutrons or protons and pions. 
The cascade particles will also decay into neutrons or protons and pions in two steps. They also have a lifetime of ($2.90 \times 10^{-10} $s). 

Next we look at the mesons, it is the pions and the kaons with lifetimes of the order of $10^{-8}$ secs that are present after the hadronization. The other mesons have significantly smaller lifetimes $10^{-17} - 10^{-23}$ secs. The kaons decay to pions and muons as well as muon neutrinos. 
As has been shown by Fromerth et. al. \cite{rafelski}, a large number of pions and  muons remain in the plasma at least till $10$ MeV. So even in the absence of inhomogeneities, a significant number density of  muons (about $10^{38}$ particles/$cm^3$) already existed in the plasma. In the presence of the inhomogeneities, the decay of the excess hyperons and kaons,  will increase the pion and muon number density of the plasma in the overdense regions. Hence after the quark hadron phase transition is complete, these overdense regions may have a higher muon concentration than the background plasma. We refer to these regions as the muon rich inhomogeneities and we will estimate the diffusion coefficients based on these parameters in Section 3. 

\subsection{Diffusion Coefficients of Nucleons}

The diffusion coefficient of nucleons have been studied in detail in both refs.\cite{ahs} and \cite{bchitre}. Both these and other references study the 
coefficients after the weak interactions have fallen out of equilibrium i.e  for temperatures less than $1$ MeV. However, overdensities are formed at around 
$200 - 100$ MeV. So diffusion of the nucleons from the overdense regions start around the same time. Since the weak interactions are in equilibrium, the protons 
and neutrons are indistinguishable at these temperatures. But the hadrons would still try to move from an overdense region and restore equilibrium in the 
baryon number distribution. The neutrons and protons would collide with the electrons and would decay into each other. Other hadrons like the hyperons and kaons will
decay and produce pions and muons. Finally, the plasma will consist of protons, neutrons, electrons, muons and their respective neutrinos. We would like to find out the 
diffusion coefficient of the nucleons at these temperatures. We will therefore concentrate on the nucleon - electron and the nucleon- muon 
cross sections. 

In a gas of lighter particles, the diffusion coefficient of a heavier particle is defined by the Einstein's equation $D = b~T$. Here, $b$ is the mobility of 
the heavier particle and $T$ is its temperature. For a Maxwellian distribution of particles, it is given by 
\begin{equation}
 b^{-1} = \frac{16\pi}{T}\int\frac{p^2dp}{3h}vp^2\sigma_te^{-E/T}
\end{equation}
Here $\sigma_t$ is the scattering cross-section, and $v$ is the velocity of the particles.

Since there are different kinds of particles in the plasma, we are dealing with multi-particle diffusion here. This depends on the concentration of the particles of different species 
in the plasma. The effective or average diffusion coefficient that is used for multi-particle diffusion is given by \cite{stewart}, 
\begin{equation}
\frac{(1 - x_i) }{D_i} = \sum_{i \neq j} \frac{x_j}{D_{ij}} 
\label{diffcoeff}
\end{equation}
Here $i$ and $j$ denote different particles of the plasma. $D_i$ denotes the diffusion coefficient of the $i^{th}$ particle and $D_{ij}$ denotes the diffusion coefficient of the $i^{th}$ particle in the presence of the  $j^{th}$ particle. Since we do not consider collision of similar particles here we have taken $i \neq j$. If $N$ be the total 
particle density, and $n_i$ be the number density of the $i^{th}$ particle, then $x_i = \frac{n_i}{N}$

We now proceed to obtain the scattering cross-sections which are required to obtain the diffusion coefficients in our case. 
The nucleon - electron cross section is dominated by form factors. However the neutron - electron and the proton - electron scattering cross-sections are not the 
same due to the presence of the electric or Mott scattering cross-section in the latter. So we have to calculate the neutron - electron and the proton - electron scattering cross-sections separately.  The diffusion coefficient due to the neutron - electron cross - section can be obtained by considering $F_1 (q^2) = 0 $ and $F_2 (q^2) = 1 $. 
Here  $F_1 (q^2)$ and $F_2 (q^2) $ are the Dirac and Pauli form factors. We do not derive the cross-section explicitly here as the derivation has been discussed in detail, previously in the literature \cite{ahs}, we only mention the relevant points.

As mentioned before we are looking at the mobility of a heavy particle passing through a gas of light particles. Here, the neutron is the heavier particle
and we assume that it is moving through a electron-positron gas.
The scattering cross section of the neutron is then given by, 
\begin{multline}
\frac{d\sigma}{d\Omega}=\frac{\alpha^2\kappa^2q^2}{16M^2E^2sin^4(\theta/2)}\frac{E'}{E}\times
\bigg[1+sin^2(\theta/2)\bigg]
\end{multline}
Here $E$ is the electron energy before the scattering and $E'$ is the electron energy after scattering and $\theta$ is the scattering angle. The transport cross section $\sigma_t$, is defined by 
\begin{equation}
\sigma_t=\int\frac{d\sigma}{d\Omega}(1-cos\theta)d\Omega
\end{equation}
Substituting the scattering cross-section we get, 
\begin{equation}
\sigma_t=3\pi\bigg[\frac{\alpha\kappa}{M}\bigg]^2
\end{equation}
Substituting the expression for the transport cross-section in the definition of the diffusion coefficient, we finally obtain,
\begin{equation}
 D_{ne} = \frac{M^2}{32 m^3} \frac{1}{\alpha \kappa^2} \frac{e^{1/T}}{Tf(T)}. 
\end{equation}
$M$, here is the neutron mass, $m$ is the electron mass, $\kappa = -1.91$ is the anomalous magnetic moment and the temperature is dimensionless as it is scaled 
by a factor of $m_e c^2$. We also have $f(T) = 1 + 3 T + 3 T^2$. 

Similar to the nucleon- electron cross-section, we also obtain the nucleon-muon scattering cross-section. The amplitude of the muon vertex is similar to the 
electron vertex. It is given by $-i e \gamma^{\nu} (q^2)$. Though the muon is heavier than the electron, it is still lighter than the neutron. Hence we can still 
consider its mass to be much smaller than the neutron mass. The heavier neutron will not move very fast compared to the lighter particles, therefore we can 
consider $q^2 \approx 0$. The form factors will then be  $F_1 = 0$ and $F_2 = 1$. The neutron vertex is given by 
$\Gamma_\mu(q^2) = \frac{i \kappa}{2 M} \sigma_{\mu\nu} q^{\nu}$. 
The differential cross-section will then be,
\begin{multline}
\frac{d\sigma}{d\Omega}=\frac{\alpha^2\kappa^2q^2}{8M^2E^2sin^4(\theta/2)}\frac{1}{1+{2Esin^2(\theta/2)}/M}\times\\
\bigg[\frac{cos^2(\theta/2)}{1-q^2/{4M^2}}\bigg(\frac{q^2}{4M^2}-1\bigg)-2sin^2(\theta/2)\bigg]\\
\end{multline}


We assume that the muon energy and mass are less than the neutron mass. This simplifies the cross-section and we can get an approximate cross-section given by, 
\begin{equation}
 \frac{d\sigma}{d\Omega} \approx K\frac{\alpha^2\kappa^2}{4M^2}[1+cosec^2(\theta/2)]
\end{equation}
Here all the constant values are put together and substituted by a single constant $K = \frac{1}{2}$. We have also assumed that the heavy neutron particle is moving through a muon-antimuon gas. The mobility of the neutron is given by the force on the neutron due to 
the medium. This force is given by the interaction cross section. Substituting in the definition of the diffusion constant, the diffusion coefficient of the 
neutron through the muon-antimuon gas is given by, 
\begin{equation}
 D_{n\mu}= \frac{M^2}{32m_\mu^3}\frac{1}{\alpha\kappa^2}\frac{e^{1/T'}}{T'f(T')}
\end{equation}
Here $T' = \frac{T}{m_\mu c^2}$.
Now that we have both $D_{ne}$ and $D_{n\mu}$, we can get the total diffusion coefficient for the neutron moving through the plasma of electrons, muons
and their anti-particles. From equation \ref{diffcoeff}, we see that it depends on the concentration of the particles in the plasma. 

We now proceed to find the diffusion coefficient of the proton moving through the electron positron gas. For proton-electron scattering, we have to take into consideration the Coulomb force. So the scattering cross section is given by, 
\begin{equation}
 \frac{d\sigma}{d\Omega} =  \frac{\alpha^2 m_e^2}{4 k^4 sin^4(\theta/2}\bigg[1+ \frac{k^2}{m_e^2}cos^2(\theta/2) \bigg]
\end{equation}
The transport cross section is then given by, 
\begin{equation}
\sigma_t = 4 \pi \alpha^2 \bigg[\frac{E_e h}{2 \pi k^2}\bigg]^2 ln (\frac{2}{\theta_0})
\end{equation}
where $\theta_0$ is the minimum scattering angle.
On substitution, we get the diffusion coefficient as, 
\begin{equation}
 D_{pe} = \frac{3 \pi}{8 \alpha^2 ln (\frac{2}{\theta_0})}\bigg[\frac{h}{2 \pi m_e} \bigg] \frac{Te^{1/T}}{f(T)}. 
\end{equation}
Similar to the proton electron cross section, we can calculate the proton muon cross section too.  The differential cross section is given by, 
\begin{multline}
\frac{d\sigma}{d\Omega}=\frac{\alpha^2}{4E^2sin^4(\theta/2)}\frac{1}{1+{2E sin^2(\theta/2)}/M} \times \\ \bigg[\bigg(1-\frac{\kappa^2q^2}{4M^2}\bigg)cos^2(\theta/2)-\frac{q^2}{2M^2}(1+\kappa)^2sin^2(\theta/2)\bigg]
\end{multline}

We are interested in the temperature dependency of the diffusion coefficient. There is no simple analytical expression for the diffusion coefficient. However we can always find them numerically by substituting the constants  and calculating the final diffusion coefficient following the same steps as before. We obtain the transport cross - section and then use it to calculate the diffusion coefficient. The values of the diffusion coefficients which are obtained numerically are given in the next section. We have not considered the collision of the neutrons and the protons here, since at these high temperatures (above 100 MeV) neutrons and protons are kept in mutual thermal equilibrium through charged-current weak interactions. This equilibrium will be maintained so long as the timescale for the weak interactions is short compared with the timescale of the cosmic expansion. Hence, we do not treat the neutron and the proton as two independent particles colliding against each other.

Immediately after the hadronization of the quarks, at thermal equilibrium, the number density of the electrons and the muons can be obtained. From ref.\cite{stuke}, we see that the number densities of leptons after hadronization is given for different values of the lepton asymmetry. We find that the number density of electrons and muon are of the same order, the authors have shown between $200 MeV - 100 MeV $  for both the cases $ \frac{n_i}{s} \approx ( 4 \times 10^5 ) MeV^{-1}$.  The number density starts to vary around $150 MeV$. If there were no inhomogeneities present then the muon number density is equal to the electron number density. However, due to the presence of the inhomogeneities, the number densities can change. Here we make an estimate in the change of the number density and then proceed to study how the diffusion coefficient changes depending on the change in the concentration
of the electrons and the muons.

\section{Diffusion Coefficients after the Quark-Hadron phase transition}

We  look at the diffusion coefficients at temperatures greater than $100$ MeV. The quark hadron phase transition occurs around $200$ MeV. The inhomogeneities are formed after the phase transition. As mentioned before, inhomogeneities with a large number of strange quarks will hadronize to give a large number 
of hyperons immediately after the phase transition. These hyperons have a short lifetime and decay into pions and muons. The pions too subsequently decay into muons. So the number density of muons would be high around these temperatures.

In ref.\cite{rafelski}, the authors have calculated the number of particles per $cm^3$ after hadronization in the absence of inhomogeneities. There we find that the number density of the electrons at $100 MeV$ is of the order of $10^{35}$ particles per $cm^3$  and the muon number density  is only slightly less than that. In their calculation, they have considered $\mu_s = \mu_d$, however in the presence of the inhomogeneities due to $Z(3)$ domain walls we have $n_s \approx 10 n_d$. This translates to a higher number of hyperons and kaons. These will decay to nucleons, electrons and the muons.

Since the plasma has the nucleons, electrons and the muons, we are using the multiparticle diffusion coefficient mentioned previously.  As $x_i$ is the fractional number density, we  have the constraint that  $\sum_{i} x_i = 1$. If all the particles are distributed evenly in the plasma then $x_{i} = 0.25$. However that is not so, hence we now need to find out what should be the values of $x_n, x_p, x_e, x_\mu $ in the inhomogeneities. Generally at these temperatures, the leptons dominate the energy density and the neutrons and protons constantly change into one another so $x_n = x_p $ and $x_n + x_p < x_e + x_\mu $. The excess of strange quarks in the inhomogeneities implies that the number density of hyperons and kaons have increased. Now as has been mentioned before, the hyperons decay into nucleons and pions. A typical decay mode of a hyperon would be $\Lambda^{0} -> p^{+} + \pi^{-} $ or $\Lambda^{0} -> n^{0} + \pi^{0}. $ The pions decay into muons ( $\pi^{-} -> \mu^{-} + \nu_{\mu} $). So the $s$ quarks would increase the number density of nucleons as well as pions and muons. The decay of the kaon may lead to two possibilities. The kaons can decay into muon/antimuon and muon neutrino or electron/positron and electron neutrino. Now as per the branching ratio of these reactions, the probability of kaons decaying into muon/antimuon and muon neutrino is far greater than the probability of kaons decaying into electron/positron and electron neutrino. According to the particle data group the branching ratio of the former is 64 \% while it is only 5\% for the latter. Thus it is clear that the inhomogeneities will give rise to an excess muon number density. For the case of the kaons decaying to  muon/antimuon and muon neutrino we use the values $x_n = x_p = 0.2$, $x_\mu = 0.4$ and $x_e = 0.2$. This case is referred to in the graphs as $x_e < x_\mu$. We have not considered the case of $x_e > x_\mu$ since the current data suggests that it is highly improbable.  The case $x_e \approx x_\mu$ is the absence of any inhomogeneity.

We first calculate the diffusion coefficients of the neutrons and the protons in a muon rich plasma and a plasma with equal number densities of electron and muons. We plot the total $D_n$ vs temperature in fig 1 and the $D_p$ vs temperature in fig 2.  The two lines denoting the two cases are as follows, the (black) dashed line denotes case (1) ($x_e < x_\mu $) and the (red) dot - dashed line denotes case (2). Now, if the particle densities depend solely on temperature (i,e in the absence of any inhomogeneities)
then between $175$ MeV and $100$ MeV the electron particle density will  be close but higher than the muon particle density \cite{husdal}.  
 As the temperature decreases, the diffusion coefficients increase. The presence of the inhomogeneities however increases the number density of the 
muons. As the number density of muons increase, we notice that the diffusion coefficient is increasing more. Thus the presence of muons changes the diffusion 
coefficient of the neutron considerably. This will definitely affect the decay of hadronic inhomogeneities at temperatures above $100$ MeV. We also plot the total $D_p$ vs temperature in fig 2.

\begin{figure}
\includegraphics[width = 80mm]{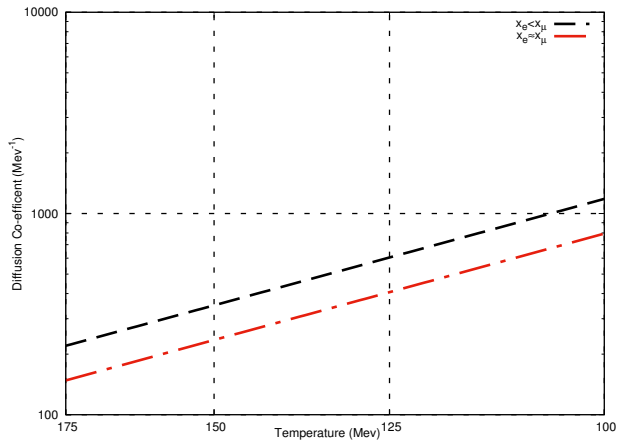}
\caption{Diffusion coefficient of neutrons in the electron, neutron and muon plasma. The (black) dashed line denotes $x_e < x_\mu$ and the (red) dot- dashed 
line denotes $x_e \approx x_\mu$. }
\end{figure}

\begin{figure}
\includegraphics[width = 80mm]{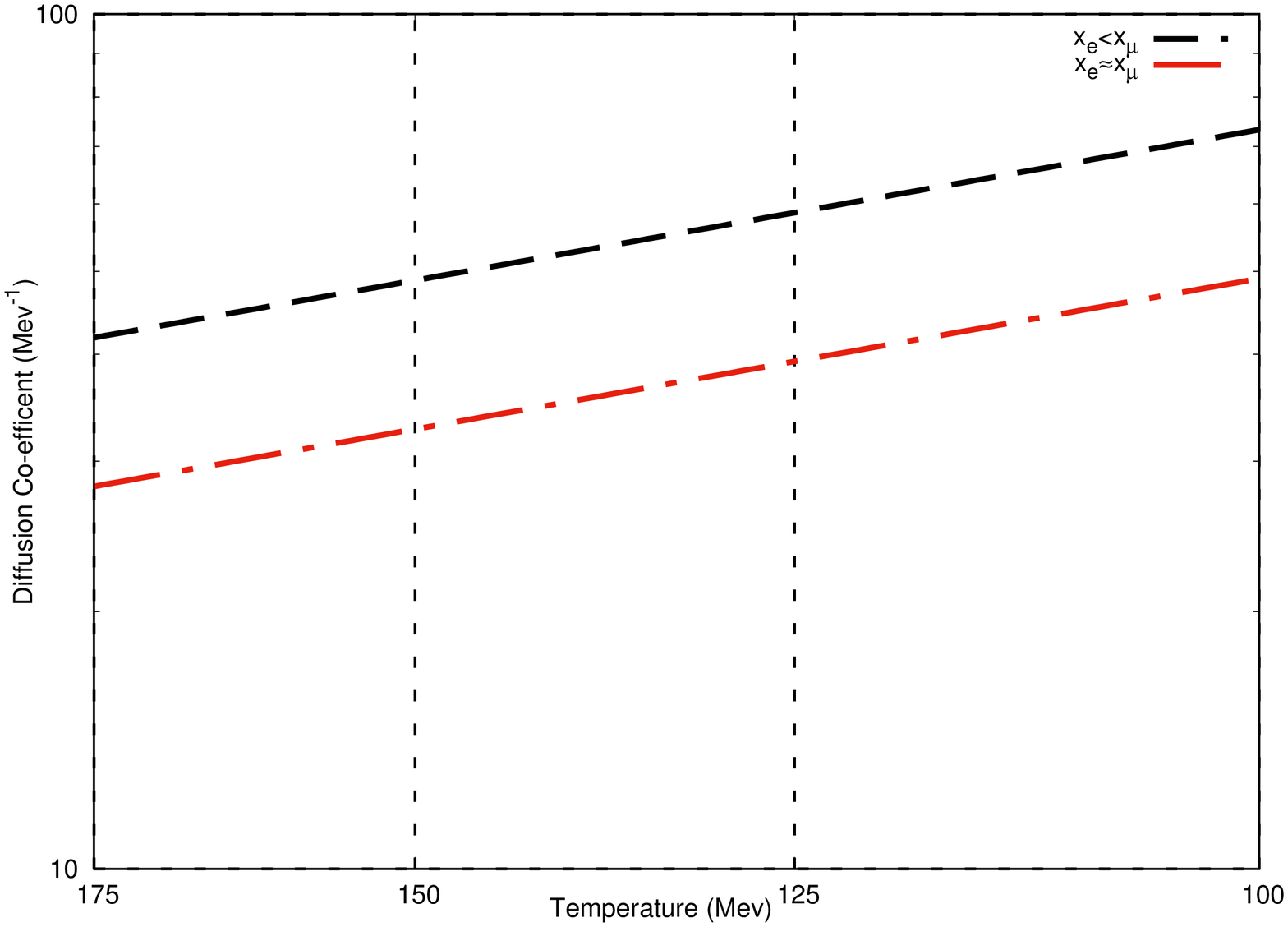}
\caption{Diffusion coefficient of protons in the electron, neutron and muon  plasma. The (black) dashed line denotes $x_e < x_\mu$ and the (red) dot- dashed 
line denotes $x_e \approx x_\mu$. }
\end{figure}

While we have calculated the number densities of the particles based on the standard decay paths and branching ratios, there is always the possibility that non-standard decays can occur and the nucleon density may be greater in the baryon inhomogeneity then in the background plasma.  
This can occur if a large number of hyperons decay via $\Lambda^{0} -> n^{0} + \pi^{0}$. The $\pi^{0}$ will decay into photons and we will have a neutron excess in the plasma. This means that the neutron density need not be fixed with respect to the muon density.  We have checked what happens if the neutron density are more but we see no significant differences.

From all the figures we can conclude that the diffusion coefficient starts to increase as the muon density is increased. Thus these graphs show that the presence of the muons changes the diffusion coefficient of the neutron/proton
through the plasma. The diffusion coefficient being increased, the nucleons move faster through the plasma. So a baryon over dense region will diffuse at a faster rate if the muon number density is higher. 
However, this only happens when temperatures are quite high. As the temperature cools
to $1$ MeV, the number density of muons go down. During this period, the contribution to the diffusion coefficient from the muons becomes negligible. Fig. 3 gives the diffusion coefficient at temperatures less than $1$ MeV.  
\begin{figure}
\includegraphics[width = 60mm, angle = 270]{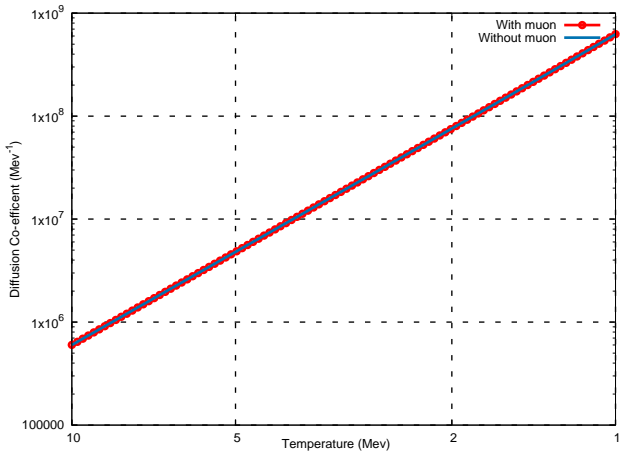}
\caption{Diffusion coefficient of neutrons in the electron, neutron, proton, and neutrino plasma at temperatures below $10$MeV. The (red) dotted line denotes the presence of muons in the 
plasma. The (blue) solid line denotes the plasma without the presence of muons. }
\end{figure}
As seen from fig. 3, the presence of the muons does not really change the diffusion coefficient around $1 $ MeV. We have thus established that the 
diffusion coefficient of the neutrons and protons change significantly due to the presence of the muons in the plasma in the overdense regions immediately after the quark hadron transition. 
We would now like to see what effect these new diffusion coefficients would have on the diffusion of hadronic inhomogeneities formed around the time of the quark hadron phase transition.

\section{Decay of inhomogeneities}
We now look at the decay of baryon inhomogeneities in the plasma around those temperatures. Baryon inhomogeneities generated at the quark hadron phase transition should be at least of the scale of $0.4$ m (at $200$ MeV) to affect nucleosynthesis \cite{kurkisuonio} calculations. So the overdensities that may affect the nucleosynthesis results will be greater than $0.5$ m. This scale is still quite small compared to the size of the universe at that time which is of the order of a few kilometers. 
Since the size of the inhomogeneity is very small compared to the size of the universe at that temperature, we can ignore the effect of the expanding universe on the decaying inhomogeneities.

We treat the inhomogeneity as a Gaussian function whose peak value at the initial time $t_0$, is given by $10^{15} MeV^3$. The average number density of the background plasma is of the order of $10^{7} MeV^3$ and baryon overdensities can be as large as $10^8$ times the background density \cite{layek}.  In general, the diffusion equation is given by, 
\begin{equation}
  \frac{D(t)}{a^2} \frac{\partial^2 n(x,t)}{\partial x^2 } = \frac{\partial n(x,t)}{\partial t }
\end{equation}
where $ D(t)$ is the diffusion coefficient which is dependent on the temperature and therefore the time in the early universe. Here $a^2$ is the scale factor of the expanding universe. Since the diffusion coefficient is time dependent, we solve the time dependent diffusion 
equation numerically to see the evolution of the inhomogeneities with time. We use a finite difference method to obtain the numerical solution of the diffusion equation for the different diffusion coefficients obtained previously. Since our diffusion coefficients are expressed in terms of temperature, we use the standard time temperature expression to obtain the diffusion equation in terms of temperature. Therefore, now our number density depends on space and temperature $n(x,T)$. We consider the inhomogeneity at $ T = 175 $ MeV, we then evolve the inhomogeneity with a given diffusion coefficient. 

\begin{figure}
\includegraphics[width = 90mm]{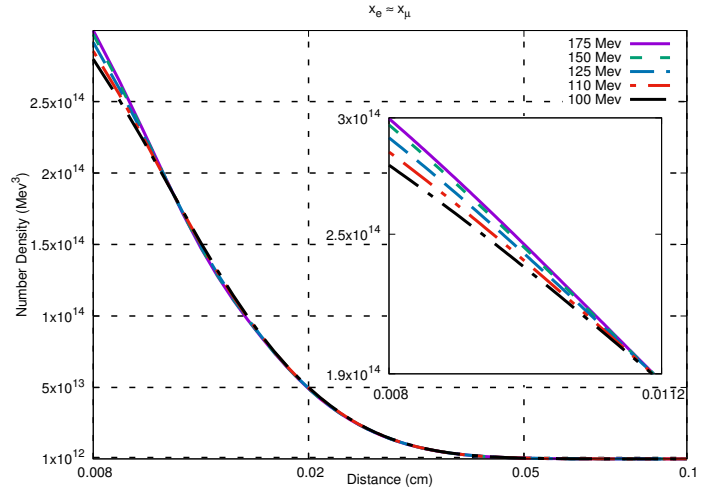}
\caption{The decay of the inhomogeneity in a plasma with equal number of electrons and muons. }
\end{figure}


 We assume for the time being that the ratio of the fractional number densities of the different particles are more or less constant through out the time of evolution of the diffusion equation. That way the diffusion coefficient is only dependent on temperature. Initially the number density decreases slowly. As time increases (temperature decreases), the peak of the inhomogeneity goes down and it spreads out in space. We initially show how an overdensity decays in a plasma which has equal numbers of electrons and muons in fig. 4, in fig. 5 we have plotted the decay of the overdensity in a muon rich plasma. 
\begin{figure}
\includegraphics[width = 90mm]{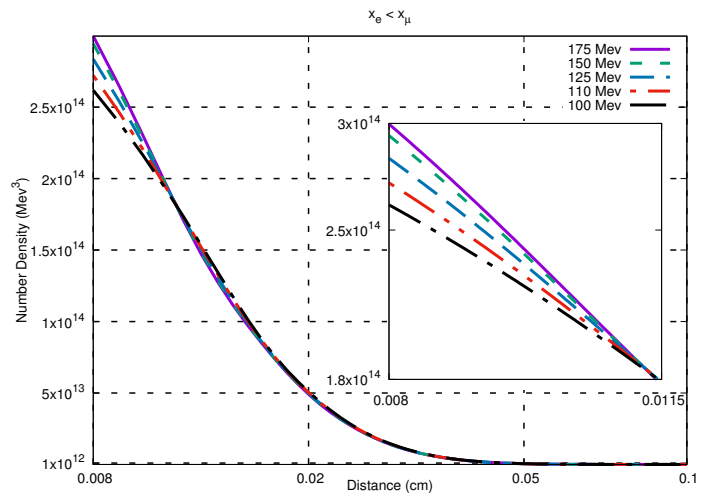}
\caption{The decay of the baryon inhomogeneity in a muon rich plasma.}
\end{figure}

From the two plots, it is clear that the muon rich inhomogeneities decay faster. The difference in the decay increases as 
the temperature cools down. The initial profile is taken to be the same at a temperature of $175$ MeV. The final profile of the inhomogeneity for the muon rich plasma is close to $2.5 \times 10^{14} MeV^3$. In the case when the electron and muon densities are the same the overdensity is close  to  about 
$2.75 \times 10^{14} MeV^3 $.  The initial size of the inhomogeneity was the same in both cases, so it indicates that the hadronic inhomogeneity decays  faster, in the presence of a large muon density. This leads us to conclude that over densities which have a larger number of strange quarks
will decay away faster after hadronization. Thus they will have little or no impact on the Big Bang Nucleosynthesis calculations.

\section{Neutrino degeneracy parameters } 
Now inhomogeneities formed due to the collapse of the Z(3) domain walls 
as mentioned before will have a larger number of strange quarks. These quarks when they hadronize will form unstable hyperons. The hyperons decay into mesons and neutrinos. Since most of them will decay due to the production of pions and pions further decay into muon and muon neutrino, there will be a larger number of muon neutrino in the plasma.

Generally, it is assumed that the three standard model neutrinos oscillate amongst themselves and have the same chemical potential at a given temperature. So in the nucleosynthesis calculations the three neutrinos are usually given the same chemical degeneracy parameter. However, it has also been shown previously, that if the lepton number densities are different for the electron neutrino and  the muon and tau neutrino, then the abundances of primordial elements are affected \cite{olive}. Therefore if Z(3) domain walls collapse and form inhomogeneities during the quark hadron phase transition, we can expect a larger number density of muon neutrinos compared to electron neutrinos. 
In nucleosynthesis calculations, the net lepton number is defined for each of the neutrinos. This is a dimensionless number defined by, 
\begin{equation}
L_i = \frac{n_{\nu_i} - {n_{\bar{\nu_i}}}} {n_\gamma}
\end{equation} 
This is related to the neutrino degeneracy parameters,   
$\xi_i = \frac{\mu_{\nu_i}}{T_\nu}$ by the equation \cite{steigman}, 
\begin{equation}
L_{\nu_i} \approx \frac{\pi^2}{12 \zeta(3)} \left( \frac{T_\nu}{T_\gamma} \right)^3 (\xi_i + \frac{\xi_i^3}{2 \pi^2} ). 
\end{equation}
During this time, the photon are slightly heated with respect to the neutrinos. $T_\nu$ is the temperature of the neutrinos and $T_\gamma$ is the temperature of the background photons. The number density can be calculated using eqn.\ref{rho} (for the detailed solution please refer to \cite{steigman}). The integral can be simplified and solved in terms of the Reimann Zeta function of order three ($\zeta(3)$). This is what determines the energy density of the neutrinos during nucleosynthesis.

We have used a standard code for the nucleosynthesis calculations. The core of the computational routines is based on Wagoner's code \cite{wagoner} but the code itself has been modified by Scott Dodelson \cite{dodelson}. The code allows us to change the neutrino degeneracies at the beginning of the calculation. The neutrino degeneracies depend on the chemical potential of the neutrinos as well as the baryon to photon ratios. The current bound on the baryon to photon ratio is quite stringent. Hence we just adhere to only one value of the baryon to photon ratio and vary only the chemical potential of the neutrinos. The chemical potentials depend on the number density and an order of magnitude estimate can be obtained from eqn. \ref{numberdensity} considering only the first term on the right hand side. The temperature is taken as a constant and the degrees of freedom are the same for all the neutrinos.

Neutrino degeneracies have been studied previously and some bounds on the degeneracy values have already been obtained \cite{olive}.  The neutrino degeneracy affects the helium and the lithium abundances more than the other abundances so we just look at the primordial helium and lithium abundances. In fig. 6, we show the abundances for $\xi_e = \xi_\mu $ in bold while we have  $\xi_e < \xi_\mu $ as the dashed line. We have considered $\eta = 3.4 \times 10^{-10} $. There are two pairs of values we have considered. One of them is $\xi_e = 0.2 $ and $\xi_\mu = 2.0 $, while the other is  $\xi_e = 0.4 $ and $\xi_\mu = 4.0 $. Our motivation for using these values are the constraints derived previously in ref.\cite{degeneracy}. Accordingly, the neutrino degeneracy parameters have to be in the ranges $-0.06 \le \xi_e  \le 1.1 $ and $|\xi_\mu| \le 6.9 $ to satisfy the CMB constraints. Our $\xi_e,\xi_\mu$ are in these ranges and the number density of the muon neutrino is about ten times that of the electron neutrino. As mentioned before, we cannot specify the decay branches of the hyperons and kaons exactly hence we tried to see what could be the maximum possible effect. Since the number density of $s$ quarks is at least $10$ times that of the $u$ and $d$ quarks hence we have calculated $\xi_\mu$ 
and $\xi_e$ by using eq.\ref{rho}. This gives us the maximum possible bound. 
 We have also tried other combinations within these parameters but none of them showed any improvement in  the final results.   

\begin{figure}
\includegraphics[width=90mm]{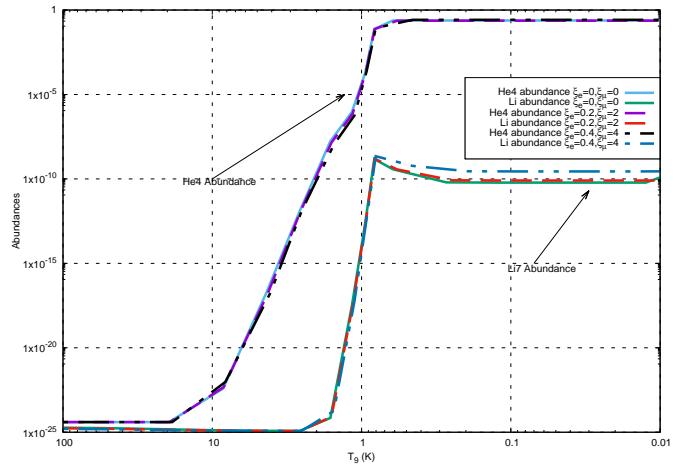}
\caption{Comparison of abundances in the presence and absence of inhomogeneities for muon degeneracy greater than the electron degeneracy.}
\end{figure}

Our results show that there are some small changes in the abundances of helium. The changes are not too significant to put constraints on the inhomogeneities. However, the lithium abundance is enhanced if we go to higher values of the degeneracies. 
Here, we have kept the muon neutrino degeneracy to be higher than the electron neutrino degeneracy at all times. Since the inhomogeneities in our model tend to decay into pions and muons, the muon neutrino degeneracy will definitely be higher than the electron neutrino degeneracy. This means that the lithium abundance will be higher than the current calculated value. As is well known, the observed lithium abundance is less than the calculated value, hence we can conclude that large inhomogeneities with a pre dominance of strange quarks will be constrained by the lithium abundance.  

Apart from the inhomogeneities from the collapsing Z(3) domains, 
there can be charged inhomogeneities too.  Charged inhomogeneities can be formed if the plasma has a small charge imbalance during the quark hadron transition \cite{raysanyal}.So we also look at the case where the electron neutrino degeneracy is greater than the muon neutrino degeneracy.This can happen if there are charged inhomogeneities in the plasma. The plot is given in fig. 7. Here however we see that both the helium abundance and the lithium abundance 
is reduced. Not only that, the large electron neutrino density also affects the neutron to proton transformation rates. Thus the beginning of the lithium production is also delayed. 
\begin{figure}
\includegraphics[width=90mm]{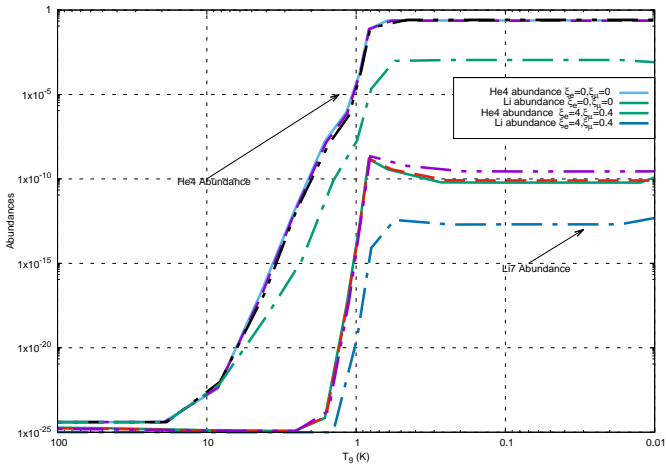}
\caption{Plot of the abundances in the presence and absence of inhomogeneities for electron neutrino degeneracy greater than the muon neutrino degeneracy included.}
\end{figure}
%

Here, we notice that when the two parameters $\xi_{\mu}$ and $\xi_{e}$ are varied there is variation in the abundances of lithium and helium. 
When  $\xi_{\mu} > \xi_{e}$, the two abundances are enhanced while if 
 $\xi_{\mu} < \xi_{e}$ the abundances are decreased. Since the decay of the inhomogeneities results in the variation of the degeneracy parameters, a detailed simulation would give us further insight in understanding the quark hadron phase transition.  
  
\section{Summary and conclusions}  
  In summary, we have shown that baryonic inhomogeneities which have a larger number of muons decay faster  compared to an inhomogeneities that are produced by a first order phase transition. Generally, in the absence of inhomogeneities, the plasma has a higher electron density compared to the muon density. In the presence of inhomogeneities however, 
the number density of muons can be increased depending on how the inhomogeneity is generated. It is quite possible that the muon density would be higher than the electron density in some of the inhomogeneities
which are formed by the collapse of $Z(3)$ domain walls. Such a scenario has never been studied before. We obtained the diffusion coefficient of the neutron and the proton in a muon rich plasma and find that at higher temperatures, it varies from the diffusion coefficient in the standard plasma. This significant change will result in the faster decay of inhomogeneities above $100$ MeV. For an inhomogeneity decaying in a plasma with equal numbers of electrons and muons, the size of the inhomogeneities need to be of the order of $0.4$ m for them to survive till the nucleosynthesis epoch. But in an muon rich plasma, the size of the inhomogeneity has to be at least 5\% bigger to survive to the nucleosynthesis epoch. So any mechanism that segregates the strange quarks more than the up and down quark must generate very large inhomogeneities to have any effect on the nucleosynthesis calculations. Inhomogeneities which have a predominance of strange quarks thus decay faster than inhomogeneities which have the different quarks in a more or less equal proportions. 

We have also looked at neutrino degeneracies generated by these inhomogeneities. Inhomogeneities which have a predominance of strange quarks will also generate a larger number of muon neutrinos compared to electron neutrinos. This means it is quite possible that a large muon neutrino degeneracy parameter is generated. We have checked whether a large muon degeneracy parameter is compatible with nucleosynthesis calculations. We find that the lithium abundance is higher than the observed lithium abundance. This puts some constraints on these inhomogeneities. Further constraints can also be obtained if a more detailed simulation of the decay of the inhomogeneity is carried out. We hope to pursue this in a later work.     
  
\section{Acknowledgement}
The authors would like to acknowledge discussions with Abhisek Saha and Soumen Nayak. SB would also like to acknowledge Payel Seth for discussions regarding the nucleosynthesis algorithms. The authors would like to thank the anonymous referee for the detailed review and critical feedback that resulted in a significant improvement of the paper.

\end{document}